\documentclass{article}

\usepackage{arxiv}

\usepackage[utf8]{inputenc} % allow utf-8 input
\usepackage[T1]{fontenc}    % use 8-bit T1 fonts
\usepackage{hyperref}       % hyperlinks
\usepackage{url}            % simple URL typesetting
\usepackage{booktabs}       % professional-quality tables
\usepackage{amsfonts}       % blackboard math symbols
\usepackage{nicefrac}       % compact symbols for 1/2, etc.
\usepackage{microtype}      % microtypography
\usepackage{cleveref}       % smart cross-referencing
\usepackage{lipsum}         % Can be removed after putting your text content
\usepackage{graphicx}
\usepackage{natbib}
\usepackage{doi}
\usepackage{comment}

\usepackage[
    type={CC},
    modifier={by},
    version={4.0},
]{doclicense}

\title{A Community Palm Model}

\date{November 6, 2024}
\usepackage{authblk}

\newbox{\orcid}\sbox{\orcid}{\includegraphics[scale=0.06]{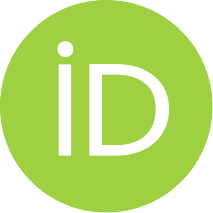}} 
\author[1]{%
	\href{https://orcid.org/0000-0002-1112-1006}{\usebox{\orcid}\hspace{1mm}Nicholas Clinton}%
}
\author[2]{%
	\href{https://orcid.org/0000-0000-0000-0000}{\usebox{\orcid}\hspace{1mm}Andreas Vollrath}%
}
\author[2]{%
	\href{https://orcid.org/0000-0000-0000-0000}{\usebox{\orcid}\hspace{1mm}Remi D'Annunzio}%
}
\author[3]{%
	\href{https://orcid.org/0000-0002-6088-5985}{\usebox{\orcid}\hspace{1mm}Desheng Liu}%
}
\author[4]{%
	\href{https://orcid.org/0000-0002-2956-530X}{\usebox{\orcid}\hspace{1mm}Henry B. Glick}%
}
\author[5]{%
	\href{https://orcid.org/0000-0003-1644-3036}{\usebox{\orcid}\hspace{1mm}Adria Descals}%
}
\author[1]{%
	\href{https://orcid.org/0000-0003-3147-9379}{\usebox{\orcid}\hspace{1mm}Oliver Guinan}%
}
\author[1]{%
	\href{https://orcid.org/0009-0005-1219-7474}{\usebox{\orcid}\hspace{1mm}Alicia Sullivan}%
}
\author[12]{%
	\href{https://orcid.org/0000-0003-2729-0901}{\usebox{\orcid}\hspace{1mm}Jacob Abramowitz}%
}
\author[7]{%
	\href{https://orcid.org/0000-0000-0000-0000}{\usebox{\orcid}\hspace{1mm}Fred Stolle}%
}
\author[8]{%
	\href{https://orcid.org/0009-0004-0099-8779}{\usebox{\orcid}\hspace{1mm}Chris Goodman}%
}
\author[1]{%
	\href{https://orcid.org/0000-0002-6504-9409}{\usebox{\orcid}\hspace{1mm}Tanya Birch}%
}
\author[1]{%
	\href{https://orcid.org/0000-0000-0000-0000}{\usebox{\orcid}\hspace{1mm}David Quinn}%
}
\author[9]{%
	\href{https://orcid.org/0000-0000-0000-0000}{\usebox{\orcid}\hspace{1mm}Olga Danylo}%
}
\author[10]{%
	\href{https://orcid.org/0000-0000-0000-0000}{\usebox{\orcid}\hspace{1mm}T. Lips}%
}
\author[1]{%
	\href{https://orcid.org/0009-0002-7791-193X}{\usebox{\orcid}\hspace{1mm}Daniel Coelho}%
}
\author[11]{%
	\href{https://orcid.org/0009-0004-4989-0355}{\usebox{\orcid}\hspace{1mm}Enikoe Bihari}%
}
\author[1]{%
	\href{https://orcid.org/0009-0005-3178-5052}{\usebox{\orcid}\hspace{1mm}Bryce Cronkite-Ratcliff}%
}
\author[11]{%
	\href{https://orcid.org/00000-0000-0000-0000}{\usebox{\orcid}\hspace{1mm}Ate Poortinga}%
}
\author[11]{%
	\href{https://orcid.org/0000-0000-0000-0000}{\usebox{\orcid}\hspace{1mm}Atena Haghighattalab}%
}
\author[12]{%
	\href{https://orcid.org/0000-0000-0000-0000}{\usebox{\orcid}\hspace{1mm}Evan Notman}%
}
\author[1]{%
	\href{https://orcid.org/00000-0000-0000-0000}{\usebox{\orcid}\hspace{1mm}Michael DeWitt}%
}
\author[1]{%
	\href{https://orcid.org/00000-0000-0000-0000}{\usebox{\orcid}\hspace{1mm}Aaron Yonas}%
}
\author[1]{%
	\href{https://orcid.org/0000-0002-3280-3858}{\usebox{\orcid}\hspace{1mm}Gennadii Donchyts}%
}
\author[1]{%
	\href{https://orcid.org/0000-0002-3280-3858}{\usebox{\orcid}\hspace{1mm}Devaja Shah}%
}
\author[11]{%
	\href{https://orcid.org/0000-0000-0000-0000}{\usebox{\orcid}\hspace{1mm}David Saah}%
}
\author[11]{%
	\href{https://orcid.org/0000-0000-0000-0000}{\usebox{\orcid}\hspace{1mm}Karis Tenneson}%
}
\author[11]{%
	\href{https://orcid.org/0000-0000-0000-0000}{\usebox{\orcid}\hspace{1mm}Nguyen Hanh Quyen}%
}
\author[4]{%
	\href{https://orcid.org/0000-0000-0000-0000}{\usebox{\orcid}\hspace{1mm}Megha Verma}%
}
\author[4]{%
	\href{https://orcid.org/0000-0000-0000-0000}{\usebox{\orcid}\hspace{1mm}Andrew Wilcox}%
}

\affil[1]{Google LLC, Mountain View, CA, USA}
\affil[2]{Food and Agriculture Organization of the United Nations, Rome, Italy}
\affil[3]{Department of Geography, The Ohio State University, Columbus, OH, USA}
\affil[4]{Unilever, Business Operations Sustainability, Englewood Cliffs, NJ, USA}
\affil[5]{CREAF, Cerdanyola del Vallès, 08193 Barcelona, Spain}
\affil[6]{University of Alabama, Huntsville, AL, USA}
\affil[7]{World Resources Institute, Washington DC, USA}
\affil[8]{NGIS, Perth, WA, Australia}
\affil[9]{RESTOR, Fraumünsterstrasse 16, 8001 Zurich, Switzerland}
\affil[10]{(personal capacity)}
\affil[11]{Spatial Informatics Group, Pleasanton, CA, USA}
\affil[11]{United States Agency for International Development, Washington DC, USA}
\affil[12]{NASA SERVIR Science Coordination Office, Marshall Space Flight Center, Huntsville, AL, USA}

\hypersetup{
pdftitle={A Community Palm Model},
pdfsubject={q-bio.NC, q-bio.QM},
}

\begin{document}
\maketitle

% Abstract
\begin{abstract}
Palm oil production has been identified as one of the major drivers of deforestation for tropical countries.  To meet supply chain objectives, soft commodity producing companies and other stakeholders need timely information of land cover dynamics in their supply shed.  However, such data are difficult to obtain from suppliers who may lack digital geographic representations of their supply sheds and production locations. There is also a proliferation of mapping products coming onto the market, which have a spectrum of methods, definitions and geographic extents that may present conflicting information and can quickly become outdated.  Here we present a “community model,” a machine learning model trained on pooled data sourced from many different stakeholders, to produce a map of palm probability at global scale. An advantage of this method is the inclusion of varied inputs, the ability to easily update the model as new training data becomes available and run the model on any year for which input imagery is available. Inclusion of diverse data sources into one probability map can help establish a shared understanding across stakeholders on the presence and absence of a land cover or commodity (in this case oil palm).  The model predictors are annual composites built from publicly available satellite imagery provided by Sentinel-1, Sentinel-2, and ALOS-2, and terrain data from Jaxa (AW3D30) and Copernicus (GLO-30).  We provide map outputs as the probability of palm in a given pixel, to reflect the model certainty of the underlying state (palm or not palm).  This version of this model provides global accuracy estimated to be ~92\% (at 0.5 probability threshold) on an independent test set.  This model, and resulting oil palm probability map products are useful for accurately identifying the geographic footprint of palm cultivation. Used in conjunction with timely deforestation information, this palm model is useful for understanding the risk of continued oil palm plantation expansion in sensitive forest areas.
\end{abstract}

% keywords can be removed
\keywords{palm \and community \and model \and remote sensing \and machine learning \and satellite \and imagery}

\section{Introduction}
Commodity supply chains have come to global attention in recent years, driven by heightened awareness of deforestation and its consequential CO\textsubscript{2} emissions resulting from land cover conversion \citep{pendrill2019deforestation, pendrill2022disentangling}. This revelation has spurred voluntary commitments from major consumer packaged goods suppliers, exemplified by Unilever, which has pledged to pursue deforestation-free practices \citep{vijay2016impacts}. Initiatives like the Round-table on Sustainable Palm Oil (RSPO) serve as proactive measures undertaken by industry stakeholders to address the environmental challenges associated with palm oil production. The urgency for accurate and timely information about commodity production locations has been further emphasized by recent regulations such as the European Union Deforestation Regulation (EUDR). Notably, the European Union identified palm oil as the primary driver of Union-driven deforestation among the eight commodities analyzed \citep{european_commission2023regulation}.

EUDR, the United Kingdom’s Environment Act, and proposed bills like the US FOREST Act, are pushing deforestation-free commodity production from voluntary commitments to a regulatory requirement for doing business in global markets.  In the case of the EUDR, this means organizations operating or placing products in the EU market must provide \citep{european_commission2023regulation}:

\begin{itemize}
  \item Geo-referenced location(s) of commodity production (EUDR Article 9, European Commission 2023).
  \item A due diligence process that demonstrates that the locations from which commodities are sourced are at low risk of deforestation (EUDR Article 10, European Commission 2023).
\end{itemize}

The risk assessment portion of the due diligence process requires large-scale, wall-to-wall, robust, and regularly updated information on land cover and land use in order to ensure that deforestation is not occurring in the sourcing locations.  Remote sensing and machine learning techniques offer promising methodologies for rapidly creating large scale land cover commodity datasets to support organizations' due diligence reporting on their supply chains.  Although commodity probability maps are not solely sufficient to ensure compliance, they are necessary to support organizations’ understandings of the impact of their supply chains, verify information provided by producers, and support due diligence reporting.

There are many land cover studies that result in high-quality maps, however they quickly go out of date and decision makers are tasked with selecting one of them, harmonizing them, or making a new one if they require maps for additional time periods, resolutions, or compliance with a particular regulation. Commercial providers also advertise commodity maps, but it can be difficult to obtain information on pricing, specifications on availability, accuracy and method of production, limiting their utility.  Recognizing the importance of open, credible, up-to-date, and consistent data products, the Forest Data Partnership has developed a model for oil palm mapping that can be easily updated with new, community-supplied information.  We consider the ability of local stakeholders to correct model output in their areas of knowledge to be an important element in the development of ethical, fair and representative models.  This communal approach involves market participants, regulated entities, regulators, researchers, and NGOs, working in collaboration to create a unified and sustainable solution.

Palm oil is a primary source of edible vegetable oil globally, with approximately 79.5 million tons produced in 2024 \citep{usda2024commodity, ritchie2021palm}.  In the period from 2000 to 2018, the conversion of forest to oil palm contributed to 7\% of global deforestation, with 29\% in Asia and 11\% in Oceania \citep{fra2020remote}.  Geographically, palm oil production is concentrated within a narrow tropical band, predominantly in three Southeast Asian countries: Indonesia, Malaysia, and Thailand \citep{ritchie2021palm}.  Indonesia alone accounts for 59\% of global production \citep{usda2024commodity}.  Consequently, the expansion of oil palm cultivation has emerged as the primary driver of deforestation in Indonesia \citep{austin2019what}. Nonetheless, the rate of oil palm expansion in Indonesia peaked around 2010 and has since decreased \citep{gaveau2022slowing}. The expansion rate of oil palm plantations in Indonesia closely mirrors the price of crude palm oil prices over time, with higher commodity prices fostering accelerated expansion \citep{gaveau2022slowing, xin2021biophysical}.

Remote sensing-based mapping of oil palm is an area of active research. A combination of optical and synthetic aperture radar (SAR) data is often employed with a classifier, such as random forest, to outperform either data source alone \citep{nomura2018more, ordway2019oil, descals2019oil, sarzynski2020combining, xu2020annual, abramowitz2023differentiating}.  \citet{li2017deep} used auto-encoders on high resolution QuickBird imagery to detect and count individual oil palm trees in Malaysia.  Working with Sentinel-1 and Sentinel-2 data, \citep{descals2021high, descals2024} used a convolutional neural network (CNN) to map oil palm globally, with distinction between industrial and smallholder plantations. \citet{danylo2021map} applied unsupervised methods to Landsat and Sentinel-1 data to map oil palm over time.  In conjunction with maps constructed from remotely sensed imagery, \citet{gaveau2018rise} and \citet{gaveau2022slowing} used manual delineation by experts to map oil palm.

The aim of this study is to leverage existing work and diverse data sources to develop a community supported Machine Learning (ML) model capable of generating annual oil palm maps from the most recent data available. Building on the existing studies mentioned above, we used the published data (where available) from these and other sources to create a communal training/validation dataset and a community ML model – a model that can be employed on-demand to create oil palm maps for time periods and locations with suitable input data.  The advantage of this type of model is that it leverages existing work, is inclusive of multiple datasets from multiple contributors, improves over time, leverages public imagery sources, doesn't lead to a proliferation of competing products and gives consumers access to the means of map production. We characterize the accuracy of the community palm model, and show that estimated probabilities output by the model can be used in statistically-based decision making frameworks.  Specifically, we use the model output at two points in time in a risk analysis framework, to quantify the risk of palm transitions in forested areas.  This risk assessment is designed to elucidate the probability of palm conversion either from, or to, other land covers in the time and area of interest.  We make the model, input data and output data available to the community for ongoing collaborative development.

\section{Methods}
\label{sec:headings}

The methods were chosen to help organizations understand the location of oil palm production, and change in such production over time as it relates to deforestation. The workflow is designed to be iterative, to facilitate rapid acquisition of additional training/validation data, model retraining and model deployment. To build this workflow, reference data were ingested into Google Earth Engine \citep{GORELICK201718} from multiple sources, and then overlaid on geographic predictors (imagery) to create training and validation sets. A TensorFlow-based neural network \citep{abadi2015tensorflow} was trained on the data and deployed to Google Vertex AI for inference at scale in Earth Engine. The model accuracy was assessed with the validation data. We performed a cross-source validation, which retrained the model leaving out one training dataset at a time to understand the contribution of each training set to the overall model performance. We also present a method to estimate risk of palm-driven deforestation by looking at predictions at two points in time for the same area.  

\subsection{Training/Testing data}

A key to this approach is the ongoing community contribution of the training and testing data that we used to build the model. Initial data sources were provided by community contributors to the Forest Data Partnership. Accuracy improves over time as additional community members contribute data to the model, which can be easily republished and re-evaluated. The data described below are provisional, and we expect stakeholders to provide more data in their areas of interest, with commensurate increase in model accuracy in those areas.  Additional data sources will be cited as new model versions are published. 

Using Maxar WorldView-3 for visual reference, Google collected 8054 1024x1024 patches at 0.3 meter resolution at nadir, independently hand-labeled by two separate annotators. Oil Palm was described to annotators as “oil palm trees planted in rows or terraces for agricultural purposes.” Bare ground patches of yet-to-be-planted palm plantations were not labeled as palm.

These were sampled to generate 326,714 points at 10 meter resolution of palm frequency in [0, 1] based on the proportion of the 10 meter pixel annotated as palm or not-palm by both annotators.  These annotations were performed on recent imagery available at the time of annotation, and were assumed to fairly represent 2021 conditions. Palm plantations operate on a growth and replanting cycle of approximately 25 years, leading to the assumption that annotation of imagery from a few years prior to 2021 would generally reflect the state of palm in 2021.

\citet{vollrath2019aglobal} provided 5,145 patches as 120x120 meter polygons representing palm in 2017.  These patches were sampled and pooled to produce 738,421 points at 10 meter resolution: palm (1) or not palm (0).

Publicly available validation data from \citet{danylo2021map} were downloaded from their GitHub site in the form of 10,303 palm presence (1) or absence (0) points (in 30x30 meter patches).  These resulted in 10,265 10-meter samples of palm (1) or not palm (0) in 2017.  

T. Lips (\textit{personal communication}) provided 180 labeled points according to palm presence or absence in 2023. These were hand-digitized using an earlier version of the model output as reference.  Specifically, points of palm presence were placed in areas of obvious under-prediction (false negatives) and points of palm absence were placed in areas of obvious over-prediction (false positives).

Publicly available data for palm plantations in Peru for 2019-2020 were obtained from \citet{fricker2022} in the form of 1288 polygons.  These data were combined with tree cover masks, a stable forest mask, the \citet{descals2021high} palm mask and subsampling (0.005 rate) at 10 meter resolution to produce 6469 palm (1) presence points.

Publicly available validation data from \citet{descals2024} were downloaded from their Zenodo site (\url{https://zenodo.org/records/13379129}) in the form of 18,812 palm presence (1) or absence (0) points (in 10x10 meter patches).  These resulted in 18,736 10-meter samples of palm (1) or not palm (0) in 2021.  These data were not used for training or model selection.

To balance the training data, pseudo-absence points (assumed not palm locations) were generated from ancillary layers on forest cover and land cover.  Specifically, Dynamic World \citep{brown2022} was used to find stable non-tree areas, defined as majority non-tree classes over a three year period (2019-2021).  Multiple forest datasets were combined to estimate stable, non-plantation, forest cover over the observation period (1984-2020).  Palm pseudo-absences were generated by stratified sampling of stable non-tree and stable forest areas in the following regions: Southeast Asia (n=14,124), Central America (n=9624), West Africa (n=35,145), and Peru (n=7819).

\subsection{Predictor data}

The predictors are annual composites built from publicly available satellite imagery provided by Sentinel-1, Sentinel-2, and ALOS-2, and terrain data from Jaxa (AW3D30) and Copernicus (GLO-30).   Specifically, the following Sentinel-2 Top-of-atmosphere (TOA) reflectance bands were used: B1, B2, B3, B4, B5, B6, B7, B8, B8A, B9, B10, B11, B12.  TOA reflectance data were used due to the increased availability of TOA data compared to data products with a higher level of processing, surface reflectance for example.   We built annual composites by taking the cloud-masked pixel- and band-wise mean for all available imagery in a calendar year.  The Cloud Score+ dataset \citep{pasquarella2023} was used for the cloud mask, with a 0.6 threshold on the 'cs\_cdf' band. The means were scaled to reflectance in approximately [0, 1] and no further normalization was performed.

Sentinel-1 data were selected from interferometric wide-swath scenes that included both VV (vertical-vertical) and VH (vertical-horizontal)  polarizations in both ascending and descending orbital paths.  Radiometric compensation for terrain using the volume scattering model described by \citet{vollrath2020angular} was used to preprocess backscattering coefficients. For a calendar year, minimum, maximum, mean, and standard deviation of backscatter were computed at each polarization and subsequently stretched to decibels.

Palsar-2 annual composites, containing HH and HV polarizations at 25 meters resolution \citep{shimada2014}, were bilinearly resampled, gap filled with the rolling 3-year mean of annual composites, and converted to scaled decibels.

Slope was derived from the ALOS AW3D30 digital surface model, a global, 30-meter surface model provided by JAXA \citep{tadono2016initial, takaku2016validation}.  Slopes were scaled to [0, 1] with no further normalization.

In a given calendar year, the eight Sentinel-1 statistics (min, max, mean, and SD for both VV and VH) were stacked with the 13 Sentinel-2 band means, the two Palsar-2 bands (HH and HV) and slope to create the annual input composites.

The palm datasets described above were combined with composites from the relevant years to produce 979,585 multi-temporal training/testing points spanning 59 countries and 10 biomes \citep{dinerstein2017ecoregion}, with class distribution:
\begin{itemize}
 \item 575,893 negatives (not palm)
 \item 532,467 positives (palm)
\end{itemize}

The pooled data were geographically partitioned into three folds based on a hexagonal decomposition of an equal-area projection, where each cell was roughly 26,000 square kilometers.  See \citet{goldblatt2018using} and \citet{fairfax2023eeager} for a detailed description of this decomposition.  Because the total boundary length of a hexagonal partition is minimized, we chose this method to minimize spatial autocorrelation between training, testing validation data.  The size and label balance of the folds vary slightly, a property we leveraged by choosing the largest fold for training.  Of the remaining two folds, one was used for model selection and hyperparameter tuning (validation) and the other for evaluation (i.e. not used in model training or selection).

\begin{figure}
	\centering
	\includegraphics[scale=0.12]{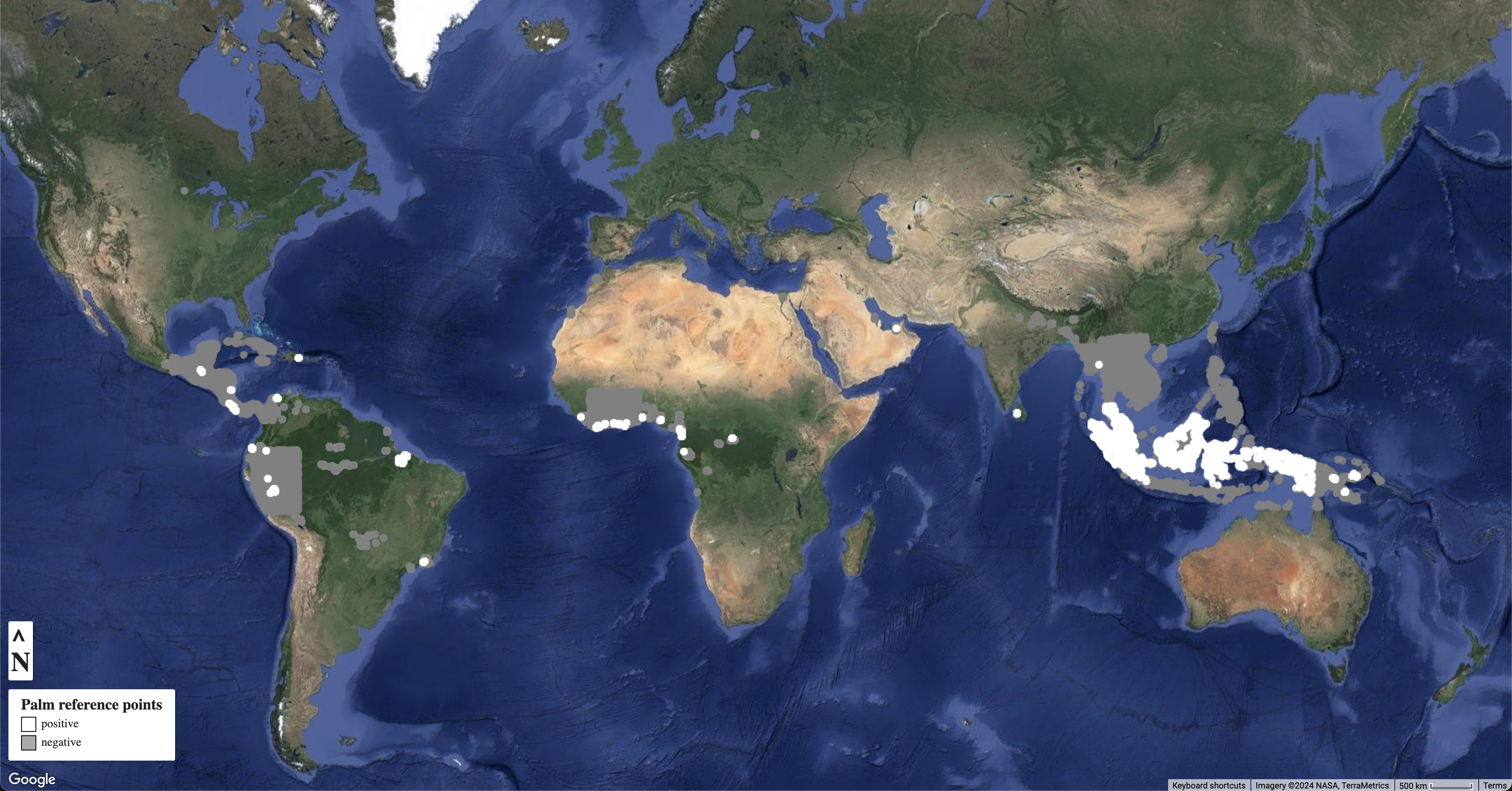}
	\caption{Location of training/testing reference locations: negatives (white), positives (gray).}
	\label{fig:fig1}
\end{figure}

\begin{figure}
	\centering
	\includegraphics[scale=0.12]{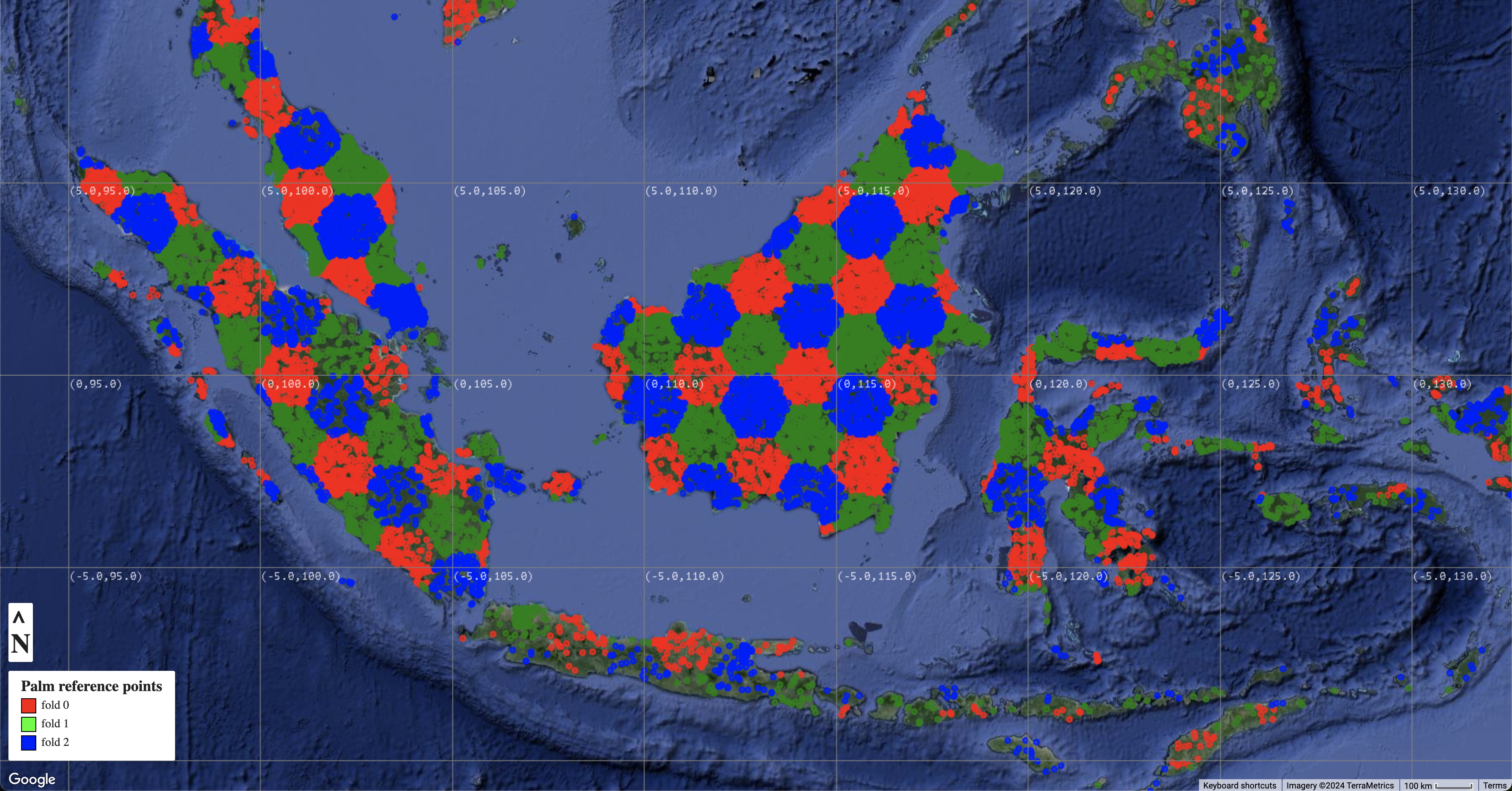}
	\caption{Distribution of folds: fold 0 (red), fold 1 (green), fold 2 (blue)}
	\label{fig:fig2}
\end{figure}

The geographic distribution of positive (palm=1) and negative (palm=0)reference points is illustrated in Figure \ref{fig:fig1}.  
The distribution of over a portion of Southeast Asia are illustrated in Figure \ref{fig:fig2}.

\subsection{Model training and inference}

The model is a neural network with four hidden layers and a single sigmoid output.  It is a per-pixel model in which the input is 1x1xC (for C bands) vector and the output (F(x\textsubscript{t})) is the estimated conditional probability of palm presence given the vector of covariates (x\textsubscript{t}) at time t.  The architecture, activation function, optimization function, learning rate, batch size, training iterations and other parameters of the model and training configuration were chosen automatically using a hyperparameter search performed on infrastructure similar to Google Cloud AutoML (\url{https://cloud.google.com/automl}).

Model inference was performed for 2020 and 2023, where the inputs to the model were determined by compositing the predictors on an annual basis.  This time frame was chosen to roughly coincide with the EUDR baseline (2020) and the most recent complete calendar year of data.

\subsection{Model accuracy assessment}

We estimate overall map accuracy in two ways.  We used the test fold (not seen in training or model selection/tuning) to compute the following accuracy measures: binary cross-entropy loss, binary overall accuracy at 0.5 threshold of output, area under the receiver operator characteristics (ROC) curve (AUC) \citep{fawcett2006introduction}, recall and precision at 0.5 threshold on the output.  As noted in \citep{zhao2014towards}, "self-evaluation" is a notorious overestimate of map accuracy.  We also use \citet{descals2024} validation data to compute the same measures (all folds and only folds 0 and 1, which are not in the same geographic fold as the training data).  The \citet{descals2024} training or validation data were not used in model training or selection.

\subsection{Risk assessment}

Here we define risk as the sum of potential land cover transitions times their associated probabilities in a region of interest (ROI).  To-palm is the transition of non-palm land cover to palm in any pixel, in units of pixel area. From-palm is the transition from palm to any non-palm land cover in any pixel, in units of pixel area. We categorize land cover into two categories: forest and non-forest, according to \citet{bourgoin2023global}.  We define the probability of the to-palm transition as $\textrm{P}(P_{t-1}=0, P_t=1)$ where P is the probability operator, and $P_t$ (italicized) is a Bernoulli distributed random variable that indicates palm presence (1) or absence (0) at time $t$.  The probability of the to-palm transition is the probability of not palm at time $t-1$ and palm at time $t$.

The pair of Bernoullis at two points in time $(P_{t-1}, P_t)$ is a bivariate Bernoulli distributed random variable \citep{marshall1985family, dai2013multivariate}.  Denote the probability of each possible state according to the contingency Table \ref{tab:table1}:

\begin{table}
	\caption{Palm indicators at two points in time}
	\centering
	\begin{tabular}{lll}
		\toprule
		\multicolumn{1}{c}{} &
		\multicolumn{2}{c}{\textit{t} = 2023} \\
		\cmidrule(r){2-3}
		\textit{t}-1 = 2020 & P\textsubscript{2023} = 1 & P\textsubscript{2023} = 1 \\
		\midrule
		P\textsubscript{2020} = 1 & \textit{p}11 = P(\textit{P}\textsubscript{t-1}=1, \textit{P}\textsubscript{t}=1) & \textit{p}10 = P(\textit{P}\textsubscript{t-1}=1, \textit{P}\textsubscript{t}=0) \\
		P\textsubscript{2020} = 0 & \textit{p}01 = P(\textit{P}\textsubscript{t-1}=0, \textit{P}\textsubscript{t}=1) & \textit{p}00 = P(\textit{P}\textsubscript{t-1}=0, \textit{P}\textsubscript{t}=0) \\
		\bottomrule
	\end{tabular}
	\label{tab:table1}
\end{table}

Note that $p11 + p10 + p01 + p00 = 1$ and that the marginal probability 

\begin{equation}
\textrm{P}(P_{2023}) = \textrm{P}(P_{2020}, P_{2023}) + \textrm{P}(\sim P_{2020}, P_{2023}) = p11 + p01
\end{equation}

where \textasciitilde\, is the negation operator, i.e. \textasciitilde\textit{P}\textsubscript{2020} == (\textit{P}\textsubscript{2020}=0).  

The objective is to obtain $p01 = \textrm{P}(\sim P_{2020} , P_{2023}) = \textrm{P}(P_{2020} = 0, P_{2023} = 1)$.  We treat the output of the model as an estimate of the parameters of the marginal Bernoulli distribution(s) in each pixel, i.e. $\textrm{F}(x_t) \approx \textrm{P}(P_t)$ for some \textit{t}.  With $\textrm{F}(x_t) \approx \textrm{P}(P_t)$ as an estimate of the marginal $\textrm{P}(P_{2023})$, we need an estimate of p11 to obtain p01 from equation 1.

Assume that the Spearman correlation $(F(x_{t-1}), F(x_t))$ in a spatio-temporal neighborhood (i.e. the correlation computed in a 310x310 meter, 31x31 pixel, neighborhood centered on each pixel in the image inputs) is a fair approximation to the Pearson correlation of the unknown indicator variables.  Note that the Pearson correlation is defined as:

\begin{equation}
\rho_{t-1,t} = (\textrm{E}[P_{t-1}*P_t] - \textrm{E}[P_{t-1}]*\textrm{E}[P_t]) / ((\textrm{Var}(P_{t-1})*\textrm{Var}(P_t))^{1/2}
\end{equation}

where E is the expectation operator and Var is the variance operator.  Note that we have estimates of the marginal expectations (e.g.  $\textrm{F}(x_t)$) and variances (e.g. $\textrm{F}(x_t)(1- \textrm{F}(x_t))$) from the model.  Also note that $\textrm{E}[P_{t-1}*P_t] = \textrm{P}(P_{t-1}=1, P_t=1) = p11$, so we can estimate p11 from the estimate of spatio-temporal correlation.  As a result, we can estimate p01 as

\begin{equation}
\textrm{p01 = P(to-palm) = P}(P_t) - \rho _{t-1,t} * (\textrm{F}(x_t)(1-\textrm{F}(x_t)) \textrm{F}(x_{t-1})(1-\textrm{F}(x_{t-1})))^1/2 +  \textrm{F}(x_t)\textrm{F}(x_{t-1})
\end{equation}

Note that we can also estimate p10 (P(from-palm)) by substituting $P_{t-1}$ for $P_t$ in equation 3.  We define the cost of a transition as the area (hectares) of a pixel $(A_{i,j})$ in which a transition occurred.  The risk of transition associated with a given ROI is therefore the sum over all $i,j$ in the ROI of A\textsubscript{i,j}*P(transition)\textsubscript{i,j}.

\section{Results}

Figures 3-6 show 2020 annual composites of the predictors in the model at a specific location. 

\begin{figure}
	\centering
	\includegraphics[scale=0.12]{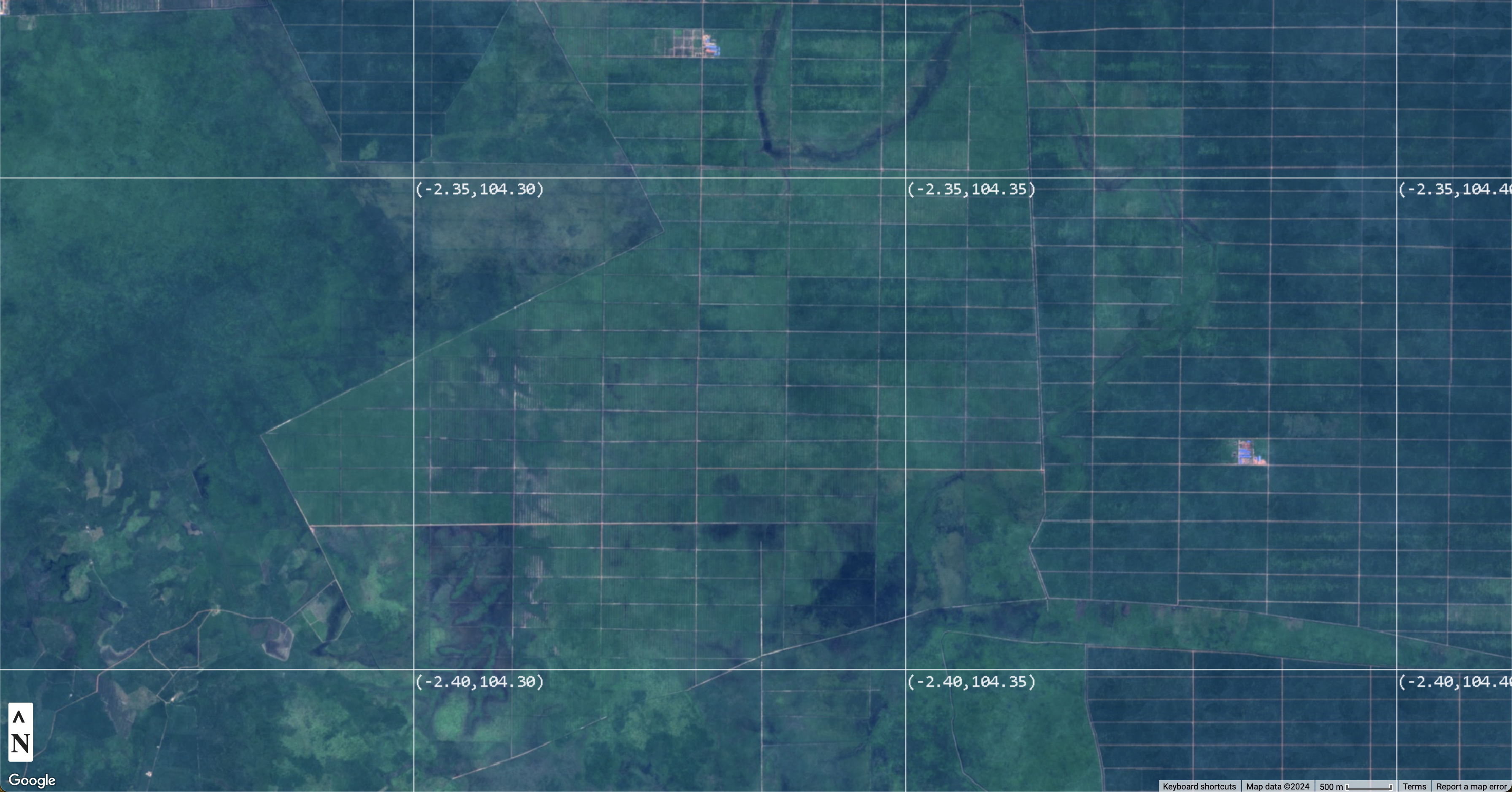}
	\caption{Annual (2020) mean composite of Sentinel-2 bands B4, B3, B2 as red, green, blue respectively.}
	\label{fig:fig3}
\end{figure}

\begin{figure}
	\centering
	\includegraphics[scale=0.12]{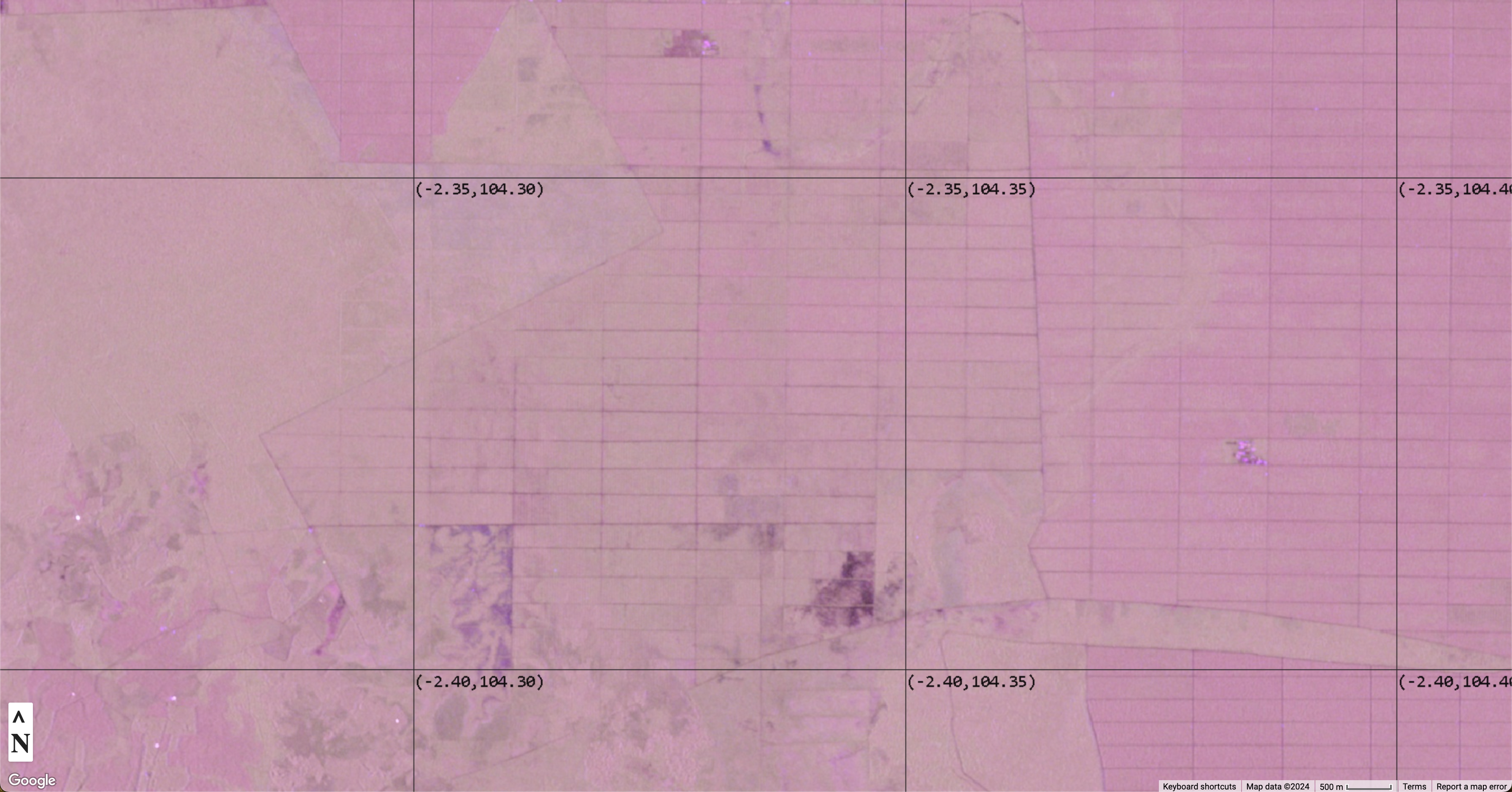}
	\caption{Annual (2020) mean composite of Sentinel-1 bands VV mean, VH mean, VV standard deviation as red, green, blue.}
	\label{fig:fig4}
\end{figure}

\begin{figure}
	\centering
	\includegraphics[scale=0.12]{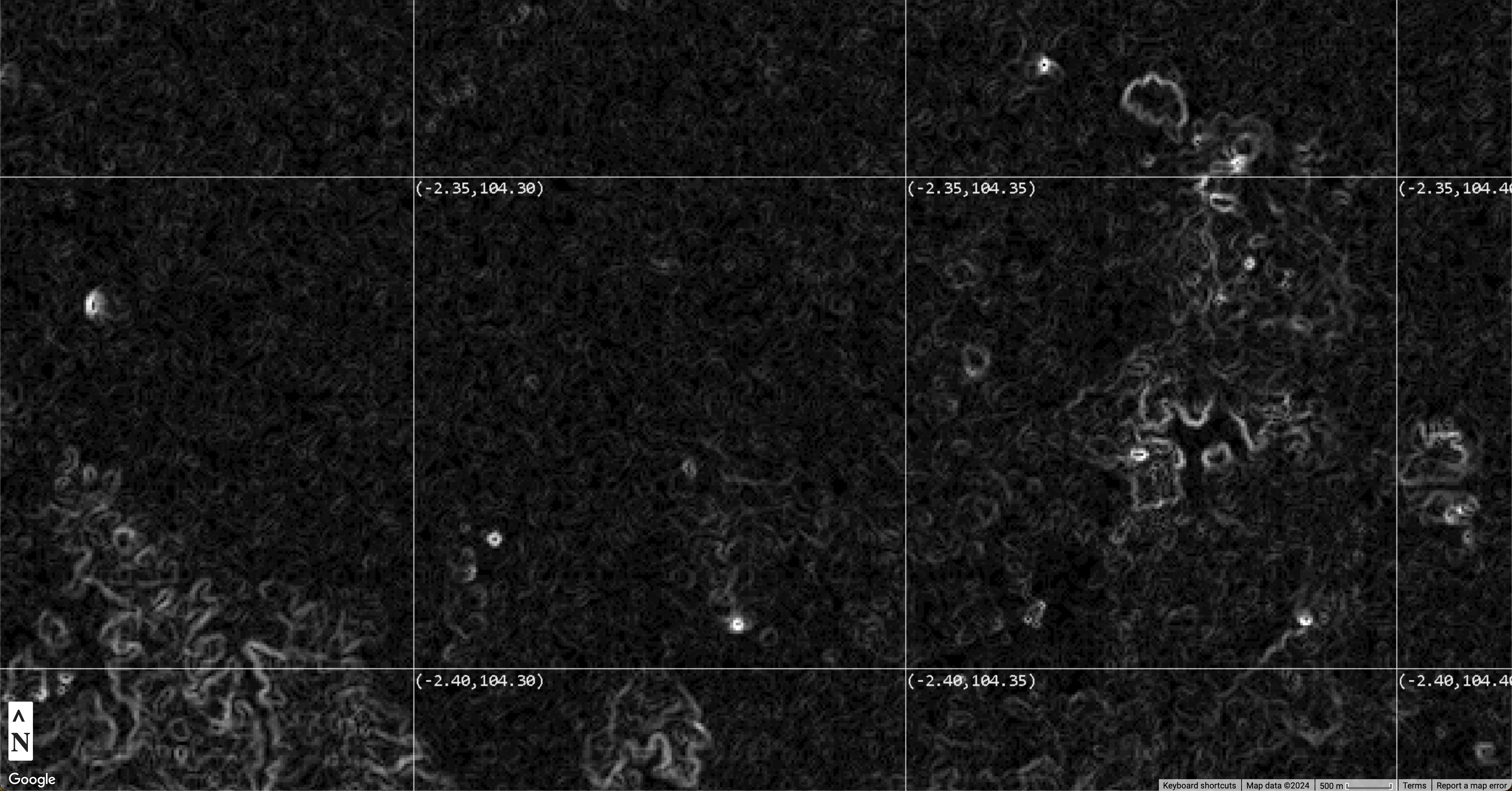}
	\caption{Slope derived from ALOS AW3D30 (low to high slope: black to white).}
	\label{fig:fig5}
\end{figure}

\begin{figure}
	\centering
	\includegraphics[scale=0.12]{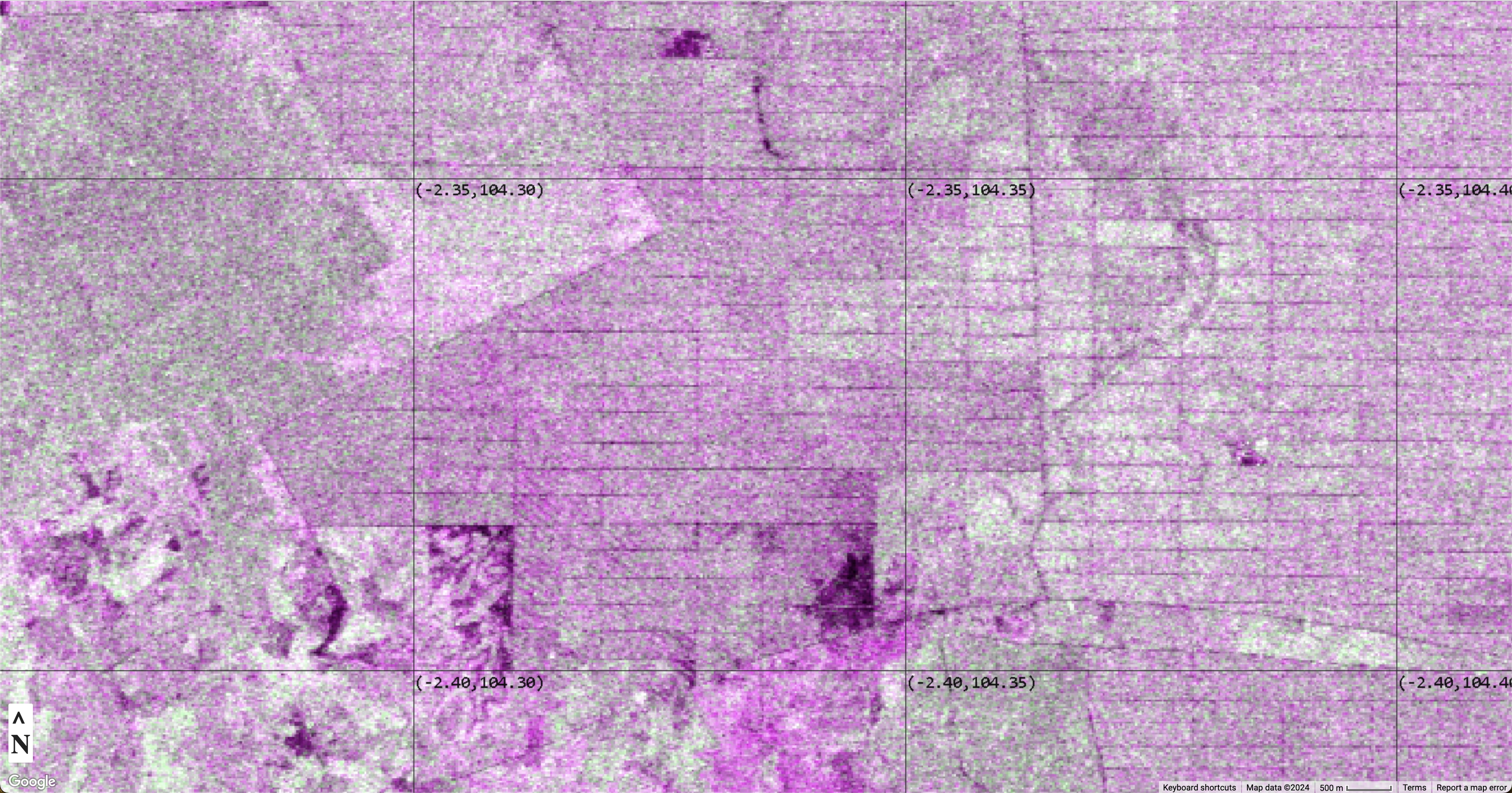}
	\caption{Annual (2020) Palsar-2 composite, HH, HV, HH bands as red, green blue.}
	\label{fig:fig6}
\end{figure}

The model is published as a trained TensorFlow model on GitHub (\url{https://github.com/google/forest-data-partnership}).  For interactive display and/or further analysis, the model can be hosted on platforms like Google Vertex AI and model output accessed from Google Earth Engine.

For palm producing countries, annual probability maps for 2020 and 2023 are produced as 10m raster image collections that are available in the Google Earth Engine public data catalog under a CC-BY 4.0 NC license for non-commercial users and commercial licensing by request (see Section \ref{modelanddata}).  Figures 6-7 shows the model output (estimated conditional probabilities) for years 2020 and 2023.

\begin{figure}
	\centering
	\includegraphics[scale=0.12]{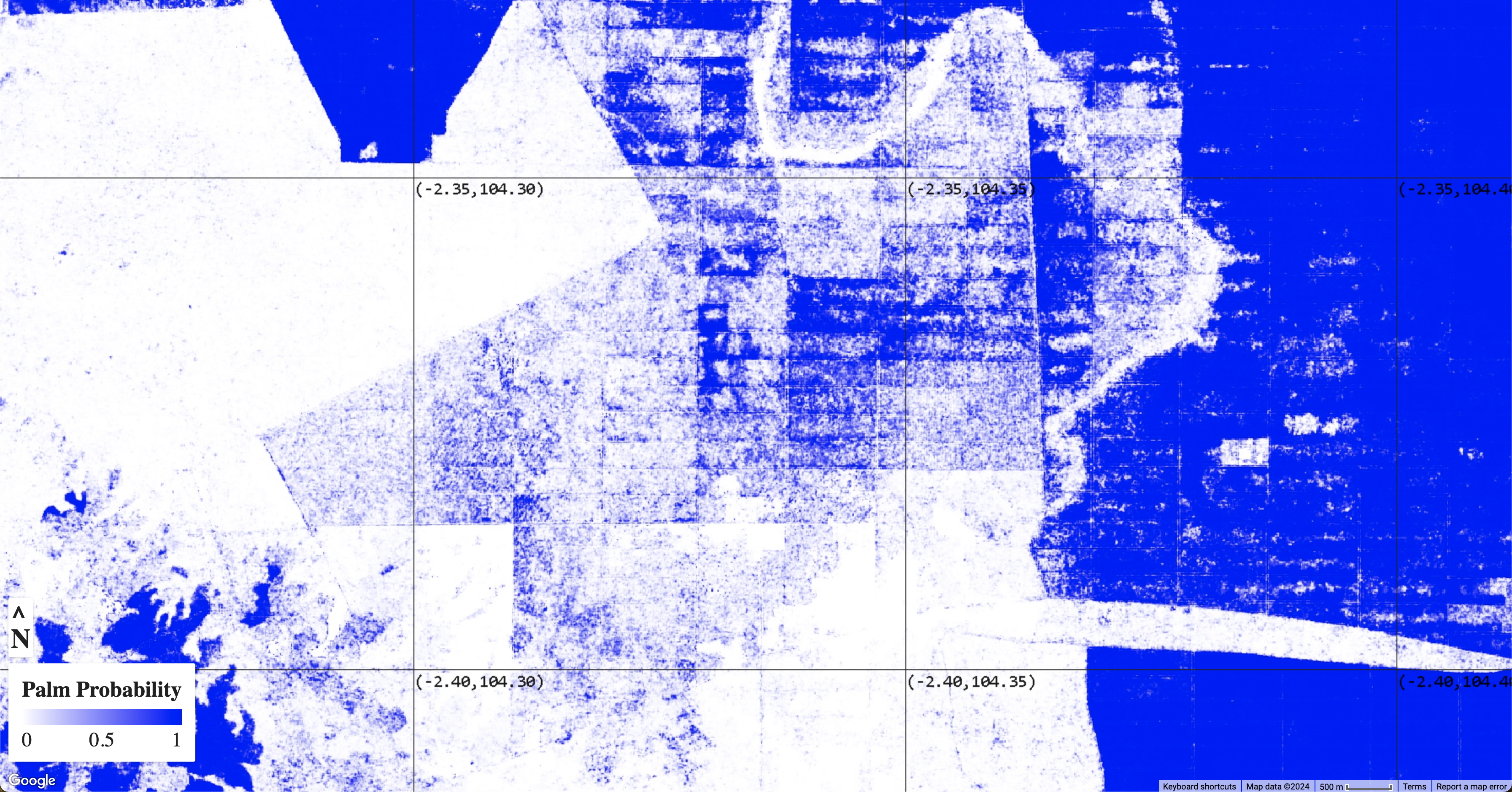}
	\caption{Estimated palm probability 2020.}
	\label{fig:fig7}
\end{figure}

\begin{figure}
	\centering
	\includegraphics[scale=0.12]{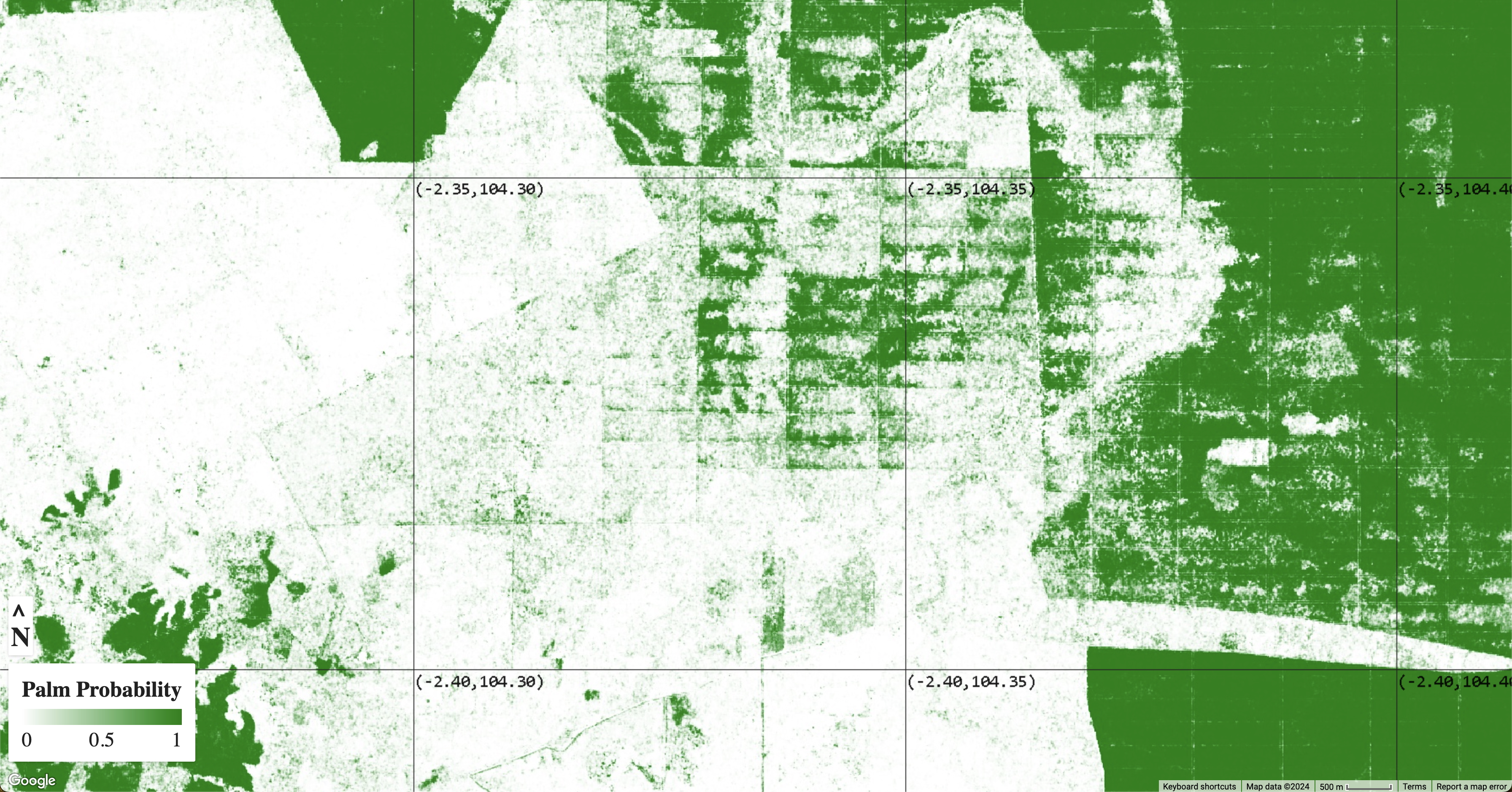}
	\caption{Estimated palm probability 2023.}
	\label{fig:fig8}
\end{figure}

\subsection{Accuracy}

Table \ref{tab:table2} shows accuracy measures computed from the geographic test fold and folds 0 and 2 of the \citet{descals2024} validation set. Binary accuracy (threshold = 0.5) is approximately 92\% on the holdout data.

\begin{table}
	\caption{Accuracy measures on multiple datasets.}
	\centering
    \begin{tabular}{ lllllll }
    \toprule
    Accuracy Measure & Test fold & Descals et al. 2024 \\
    \midrule
    Cross-entropy loss & 0.097 & 0.183 \\
    Binary accuracy @ 0.5 threshold & 0.969 & 0.924 \\
    Area under the ROC curve (AUC) & 0.992 & 0.976 \\
    Recall @ 0.5 threshold & 0.968 & 0.913 \\
    Precision @ 0.5 threshold & 0.969 & 0.600 \\
    \end{tabular}
	\label{tab:table2}
\end{table}

\begin{figure}
	\centering
	\includegraphics[scale=0.2]{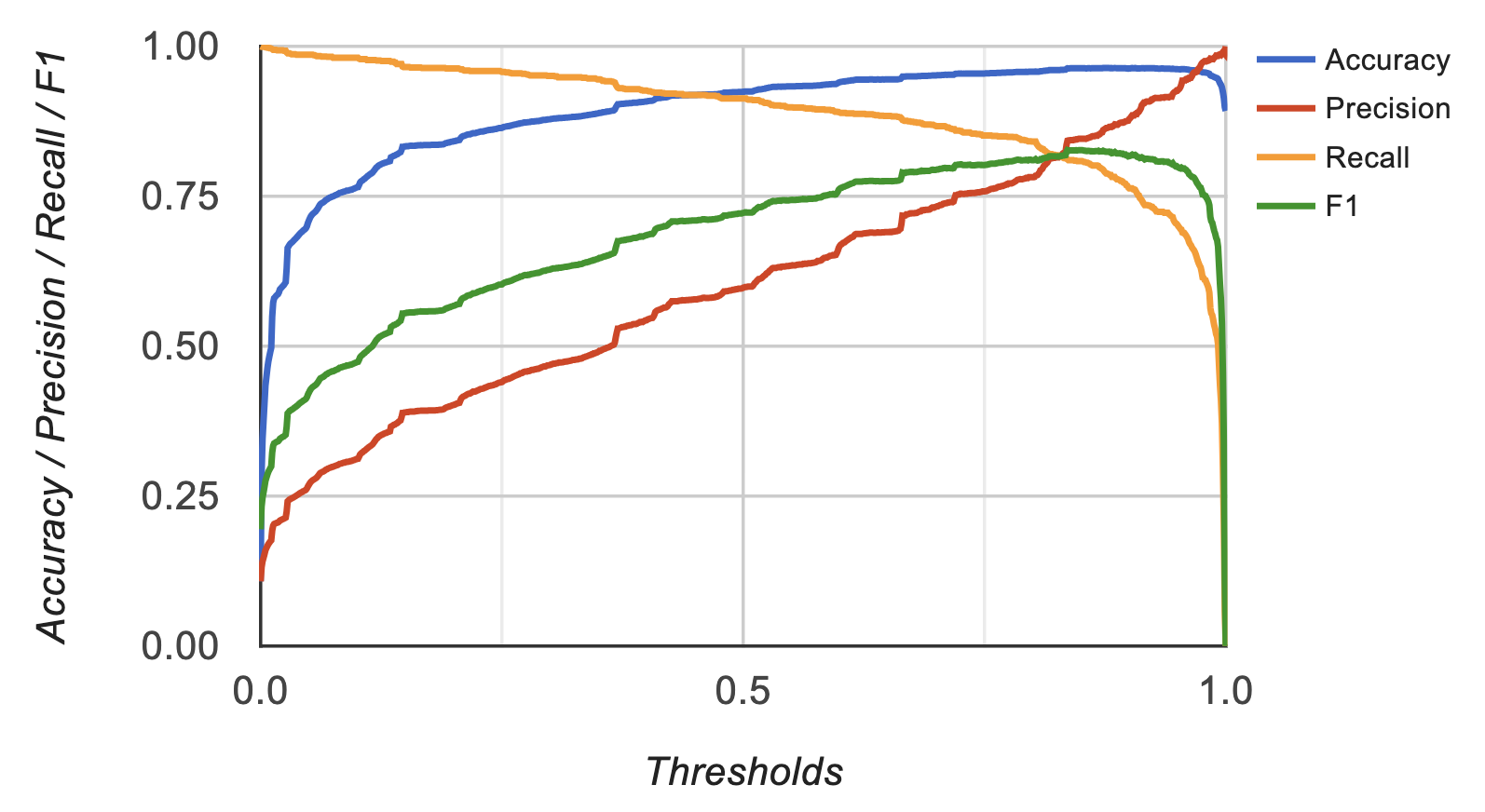}
	\caption{Accuracy, F1, precision and recall curves on Descals et al. 2024 validation data.}
	\label{fig:fig9}
\end{figure}

The accuracy assessment illustrates several important issues.  The test fold consistently shows higher accuracy than the \citet{descals2024} validation set, suggesting potential self-evaluation bias and/or dependence between folds.  Since the \citet{descals2024} validation data were not seen in model training or selection, and were restricted to geographic locations different from the training data, these are a more conservative estimate of overall model performance.  Note that the precision at threshold 0.5 is ~60\% on this set, suggesting overestimation in non-palm areas at low thresholds.  Figure \ref{fig:fig9} shows overall accuracy, F1, precision and recall curves for the model on the \citet{descals2024} validation set.  Note that the threshold which optimizes F1 score is closer to 0.82, where precision and recall are ~82\%.  Figure \ref{fig:fig9} also illustrates the extreme imbalance (towards negatives) of the the \citet{descals2024} validation set.  Specifically, overall accuracy increases with increasing threshold, which is near 96\% at the 0.82 threshold.

\subsection{Risk assessment}

\begin{figure}
	\centering
	\includegraphics[scale=0.12]{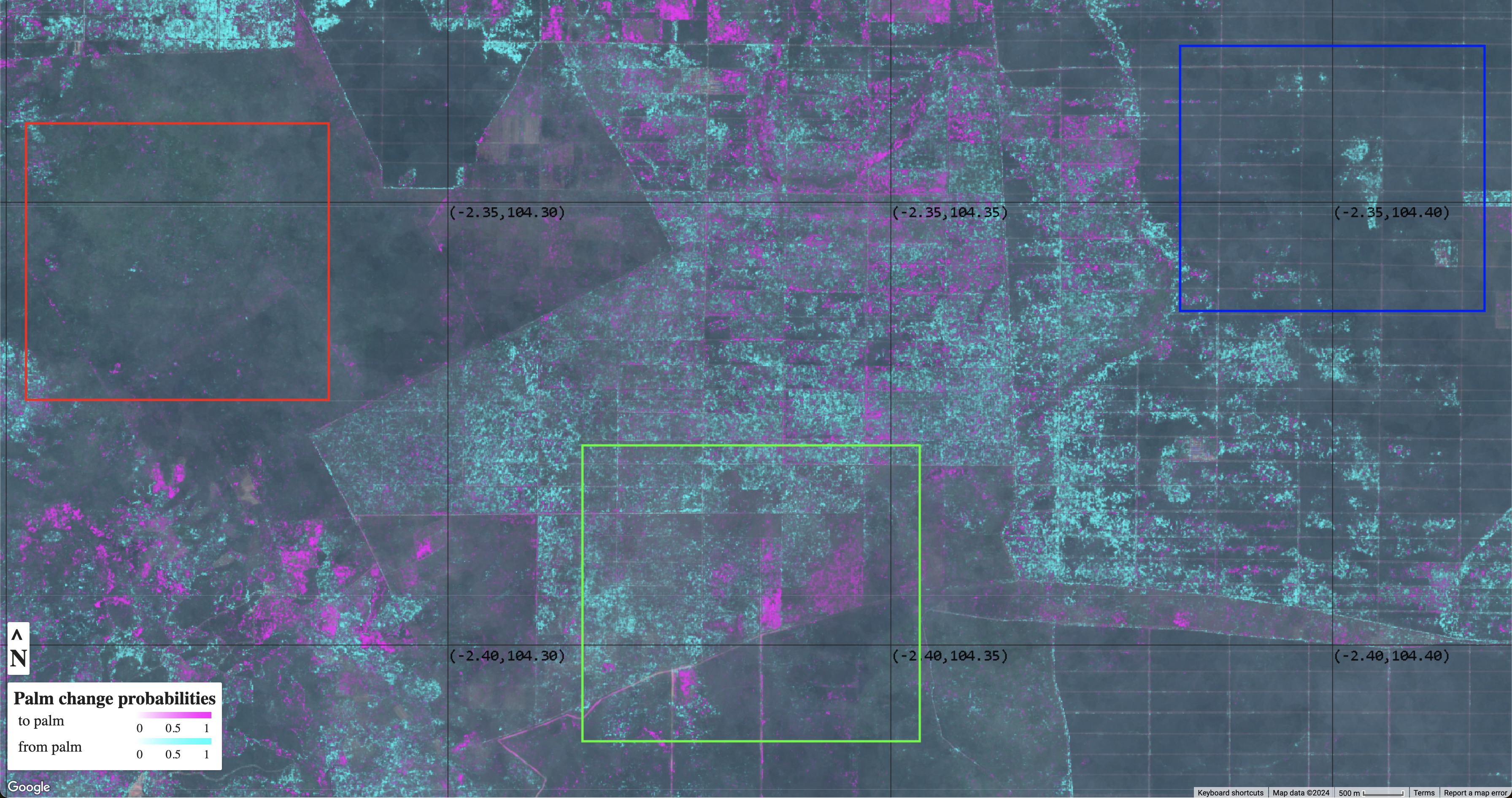}
	\caption{Hypothetical sourcing domains and palm transition probabilities, with 2023 Sentinel-2 RGB background.}
	\label{fig:fig10}
\end{figure}

Figure \ref{fig:fig10} shows the estimated probability of palm transitions from years 2020 to 2023, computed according to equation 3 or its analog for $P_{t-1}$.  The three regions shown in Fig. \ref{fig:fig10} correspond to hypothetical sourcing domains.  Inspection of the imagery suggests that region 1 (red) represents an area without palm, the region 2 (green) represents a mixed area, and region 3 (blue) represents an established palm plantation.  The risk of palm transition associated with each ROI was computed as the sum of estimated probability of the transition in each pixel times cost of the transition (pixel area in hectares).  The results are shown in Table \ref{tab:table3}.

\begin{table}
	\caption{Transition risk in forest categories.}
	\centering
    \begin{tabular}{ lrrr }
    \toprule
    Areas (ha) & Region 1 & Region 2 & Region 3 \\
    \midrule
    \textbf{Forest} & 983 & 102 & 0 \\
    To-palm risk (forest) & 17.9 & 2.4 & N/A \\
    From-palm risk (forest) & 14.9 & 1.7 & N/A \\
    \textbf{Non-forest} & 331 & 1465 & 1266 \\
    To-palm risk (non-forest) & 5.1 & 83.6 & 4.8 \\
    From-palm risk (non-forest) & 5.3 & 126.5 & 32.9 \\
    \end{tabular}
	\label{tab:table3}
\end{table}

\section{Discussion}

The method described here leverages existing datasets to create a model applicable to any time or place given sufficient input data.  This is important in the context of regulatory processes that require harmonization of potentially ambiguous information to demonstrate sustainability.  A community model is useful for a logical, data-driven approach to model probability of existence of a specific commodity at a certain location. This approach is of ever increasing importance due to the proliferation of maps and the need for companies to demonstrate compliance with existing and emerging nature protection laws.  Specifically, regulations intended to ensure commodities are not sourced from recently deforested places require the disambiguation of conversion of natural forest to commodity production from other kinds of forest management.

Interpreting the model output as probabilities retains local uncertainty in the underlying condition, which is unknown. The global oil palm probability map developed with this method can be a two-class (palm versus not-palm) classification by thresholding probabilities using automatic methods such as Otsu segmentation \citep{otsu1979threshold}.  The threshold determined by Otsu for the 2023 palm map (roughly the area of Figure 10) is 0.506.  Thresholding the 2023 palm in Figure \ref{fig:fig7} at 0.506 indicates 5842 hectares of planted palm in the area of Figure \ref{fig:fig10}.  From Table \ref{tab:table2}, recall is relatively high at a 0.5 threshold, but precision is low].  Any selected threshold represents a trade-off between recall and precision as shown in Figure \ref{fig:fig9}.  For this reason, it is incumbent on the user of the data to determine a threshold suitable for their use case and region of interest, for example using Otsu segmentation to select an optimal threshold for local conditions. However, threshold delineation is not required, as the probabilities can be used directly. Treating the probabilities as an estimated distribution over the unknown (and assumed hidden) Bernoulli random variable (Palm: {0, 1}), indicates 6317 hectares of planted palm in Figure \ref{fig:fig7} – an ~8\% difference compared to the threshold method.

\begin{figure}
	\centering
	\includegraphics[scale=0.12]{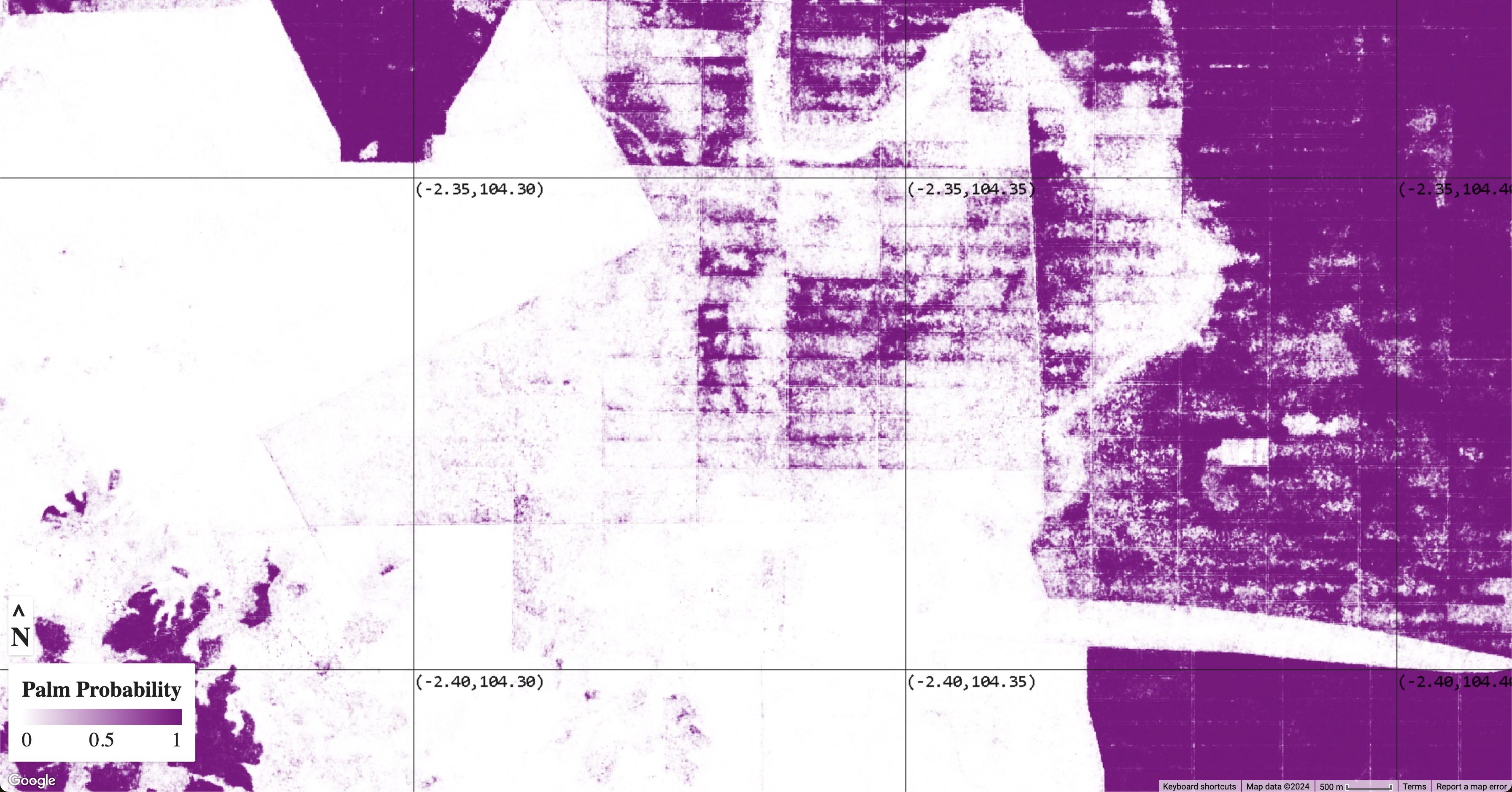}
	\caption{Estimated palm probability: both 2020 and 2023.}
	\label{fig:fig11}
\end{figure}

Since the contingency table can be specified completely with the assumptions described previously, it can be used to generate estimates of the probability of different scenarios while accounting for uncertainty in model estimates at multiple points in time.  For example, the estimate of p11 is the probability of "stable" palm, or places that were palm in both 2020 and 2023 (Figure \ref{fig:fig11}).  Estimates of persistent palm seem less noisy than single year output, but additional testing is required to confirm that.  Risk can be computed for any scenario, in units of area, using the method described.  Although risks computed in this manner can be used directly in supply chain analysis, due to potential model error, the risks should be interpreted with caution.  Specifically, these data can indicate the need for additional verification or characterization of high risk areas, but should be combined with other evidence to establish risk definitively.

For example, consider the three regions in Figure \ref{fig:fig10} and Table \ref{tab:table3}. In region 1, the majority of the area is in forest, as defined by \citet{bourgoin2023global}.  Although the to-palm risk is ~18 hectares, there is also from-palm risk of ~15 hectares.  In region 1, we conclude that the model output is noisy, as the from-palm transition risk roughly balances the to-palm transition risk.  In region 2, there is some forest, and small amounts of transition risk, which we conclude to be either noise or active palm plantation management.  Region 3 is all non-forest.  In region 3, there are relatively high amounts of transition, which we assume to be palm management.  We don't interpret any of these areas as particularly high risk.  Limitations of this approach are discussed below.

For such a model to be effective in meeting sustainability commitments, oil palm maps must be produced on a yearly basis, since compliance is determined annually.  The model also needs to be easy to update, meaning that stakeholders can contribute data in their areas of interest, if the model output is not representative of that area.  This is especially important from an ethical perspective, where local stakeholders need inclusion in a process that could affect their livelihoods.  Further verification of submissions may be required where submissions are found to reduce overall or local model accuracy.  A single iteration of the process consists of an examination of model results by a relevant stakeholder, creation of additional training/validation data in the stakeholder's ROI, model refinement and retraining using the new data, and publication of the updated model.  That workflow was demonstrated here.  Specifically, preliminary model results were provided to T. Lips, who created a new training dataset, after which point the improved model was retrained and republished.  Additional iterations and version tracking are needed to keep the model inclusive, accurate and up-to-date.

\subsection{Limitations}

Although uncertainty is implicit in the model output, which is expressed as a probability, the model could be biased due to training data biases or an inability to generalize the training data in the model weights.  This bias could be reduced through enhancements to model architecture, for example through inputting spatio-temporal patches of inputs instead of single pixel vectors from annual composites.  There could also be high variance in the model output.  This is illustrated by the analysis of region 1, in which noise in model output is assumed to account for a substantial amount of from- and to-palm transitions in the region.  Variance could be reduced through collection of a more geographically representative and balanced training data set.  We suspect variance is also reduced by combining multiple years of predictions to isolate stable crops of interest (e.g. Figure \ref{fig:fig11}), but additional research is needed to confirm that suspicion.  Presently, known limitations resulting from model bias and/or variance mean quality of predictions will vary geographically and results should be interpreted with caution.

The probabilistic framework on which the risk assessment is based relies on strong assumptions.  We assume that the true state of a pixel can be adequately represented by a Bernoulli random variable, implying that a pixel can be in only one of two states: palm or not palm.  We make little effort to determine, or even to define, such states, which are simply taught to the model directly from the data.  Given a vector of predictors, we assume the model output is a suitable representation of the conditional probability of oil palm at a given location, and treat it as the marginal probability of the unknown indicator variable.  Finally, the estimate of palm conversion depends on the estimation of correlation of the unknown indicators from the Spearman rank correlation of the model outputs.  The Spearman correlation is probably an overestimate of the Pearson correlation between the unknown indicators, since the rank-correlation can be high for non-linearly related variables.  This would overestimate p11 and consequently underestimate the transition risk.

\subsection{Future Work}

We intend for the model presented here to continuously improve.  To maximize effectiveness, the model should evolve as relevant community stakeholders contribute additional data.  To this end, we invite the larger community of local, regional, and global stakeholders to submit informed feedback, additional training data and/or other information (feedback form: \url{https://goo.gle/fdap-data}.  Improvements are not limited to a steadily increasing amount of training and/or validation data.  The model could be improved through changes in input format (e.g. patches, which contain more spatial information) and/or model architecture.  The inputs could be improved through incorporation of more information as new sensors come online or through learned embedding spaces (\citep{jean2019tile2vec}).  Community contributions to model and or input data are welcome through our GitHub repository (see Section \ref{modelanddata}).  We expect modeling other land covers of interest (e.g. cocoa, coffee, rubber) in a community framework to increase accuracy and availability of information about the spatio-temporal distribution of commodities and relationship to deforestation.

\section{Conclusion}

We have presented a “community model” approach to generate open, timely, consistent and global land cover maps for oil palm production. The strength of this approach comes from cross stakeholder data pooling for training, ease of incorporating additional training data over time and the open availability of the trained machine learning models and probability maps. The community land cover model we present here can be operated at multiple points in time and/or multiple geographic locations and can take multiple diverse inputs.

This model, and resulting oil palm probability maps are useful for accurately identifying the geographic footprint of oil palm cultivation, and how that footprint is changing over time. We have demonstrated how the analysis of the model outputs at multiple points in time can be used to infer oil palm dynamics. We recognize that the commodity probability maps are not solely sufficient to quantify palm oil driven deforestation, however they are necessary to disambiguate deforestation of natural forests from other kinds of land cover transitions.  Used in conjunction with other geospatial information, such as global forest cover, deforestation alerts and high resolution imagery, this community palm oil map and risk indicators will support organizations’ understanding of the impact of their supply chains, help to verify information provided by producers, and support due diligence reporting.  Specifically for regulatory mechanisms like the EUDR, we intend for the community model and resulting data products to support the  “convergence of evidence” approach developed by the Forest Data Partnership for due diligence statements and compliance claims.

Ongoing research and development is needed to continue to refine the community palm model and map products. Additionally, expansion of the community model method to commodities beyond palm is needed to sufficiently characterize how commodity production is changing over time and impacting the world’s remaining forests.

\section{Model and Data} \label{modelanddata}

Interactive palm oil probability viewer:
\begin{center}
\url{https://forestdatapartnership.projects.earthengine.app/view/palm}
\end{center}

Trained models and model hosting example:
\begin{center}
\url{https://github.com/google/forest-data-partnership/tree/main}
\end{center}

Earth Engine asset:
\\
\texttt{ee.ImageCollection("projects/forestdatapartnership/assets/palm/model\_2024a")}

\section{Acknowledgements}

We would like to acknowledge USAID funding of the Forest Data Partnership.  Matt Hancher provided a valuable review.

\bibliographystyle{unsrtnat}
\bibliography{palm_paper}  

\end{document}